\documentclass[journal]{IEEEtran}
\usepackage{graphicx}
\usepackage{float}
\usepackage{amsmath}
\usepackage{pgfplots}
\pgfplotsset{compat=newest}
\usepackage{subfig}
\usepackage{hyperref}
\usepackage{tikz}
\usetikzlibrary{arrows.meta, positioning}

\title{ART-Rx: A Proportional-Integral-Derivative (PID) Controlled Adaptive Real-Time Threshold Receiver for Molecular Communication}
\author{Hongbin Ni, \textit{Student Member, IEEE} and Ozgur B. Akan, \textit{Fellow, IEEE}\thanks{The authors are with the Internet of Everything Group, Electrical Engineering Division, Department of Engineering, University of Cambridge, CB3 0FA Cambridge, U.K. (e-mail: hn345@cam.ac.uk).\\ Ozgur B. Akan is also with the Center for neXt-Generation Communications (CXC), Department of Electrical and Electronics Engineering, Koc University, 34450 Istanbul, Turkey (e-mail: oba21@cam.ac.uk; akan@ku.edu.tr).\\ }}

\begin{document}

\maketitle

\begin{abstract}
Molecular communication (MC) in microfluidic channels faces significant challenges in signal detection due to the stochastic nature of molecule propagation and dynamic, noisy environments. Conventional detection methods often struggle under varying channel conditions, leading to high bit error rates (BER) and reduced communication efficiency. This paper introduces ART-Rx, a novel Adaptive Real-Time Threshold Receiver for MC that addresses these challenges. Implemented within a conceptual system-on-chip (SoC), ART-Rx employs a Proportional-Integral-Derivative (PID) controller to dynamically adjust the detection threshold based on observed errors in real time. Comprehensive simulations using MATLAB and Smoldyn compare ART-Rx's performance against a statistically optimal detection threshold across various scenarios, including different levels of interference, concentration shift keying (CSK) levels, flow velocities, transmitter-receiver distances, diffusion coefficients, and binding rates. The results demonstrate that ART-Rx significantly outperforms conventional methods, maintaining consistently low BER and bit error probabilities (BEP) even in high-noise conditions and extreme channel environments. The system exhibits exceptional robustness to interference and shows the potential to enable higher data rates in CSK modulation. Furthermore, because ART-Rx is effectively adaptable to varying environmental conditions in microfluidic channels, it offers a computationally efficient and straightforward approach to enhance signal detection in nanoscale communication systems. This approach presents a promising control theory-based solution to improve the reliability of data transmission in practical MC systems, with potential applications in healthcare, brain-machine interfaces (BMI), and the Internet of Bio-Nano Things (IoBNT).
\end{abstract}


\section{Introduction}

\IEEEPARstart{M}{olecular} communication (MC) or particle-based communication, a paradigm inspired by nature, has emerged as a promising solution to communicating with biological organisms where traditional communication methods have been shown to be ineffective \cite{Nakano2013MolecularCommunication, Xiao2024WhatCarriers, Nakano2014MolecularIssues, Farsad2016ACommunication}. By leveraging biochemical mechanisms for the transmission of information, MC is also believed to play a vital role in the realization of the Internet of Everything (IoE) \cite{Akan2023InternetUniverse}, particularly the Internet of Bio-Nano Things (IoBNT) \cite{MuratKuscu2021InternetChallenges, Akyildiz2015TheThings}, and the concept of digital twins through the extension of connectivity to nanoscale and biological environments \cite{Chude-Okonkwo2021ConceptualImplementation}. This unconventional bio-inspired technique encodes information with one or more types of information molecules (IM) at the transmitter end, which are then propagated to a receiver through various mechanisms such as channel diffusion \cite{Pierobon2014FundamentalsNanonetworks, Wang2017DiffusionChallenges, Kuscu2016OnBiosensors, Kuscu2016ModelingReceiver, Kuscu2019TransmitterTechniques}, mimicking methods of communication commonly found in the natural world. As research in bioengineering and nanotechnology continues to advance, there are vast opportunities \cite{Nakano2012MolecularChallenges} in the realm of MC for the development of bio-nanoscale communication systems that could potentially revolutionize fields such as healthcare \cite{Kuscu2016OnBiosensors, Akyildiz2008Nanonetworks:Paradigm, Veletic2020ModelingDelivery}, nanomachines \cite{Nakano2014MolecularIssues, Kuscu2016ModelingReceiver, Veletic2020ModelingDelivery}, and brain machine interfaces (BMI) \cite{Nakano2014MolecularIssues, Nakano2012MolecularChallenges}.

Microfluidic channel-based MC systems \cite{Kuscu2016OnBiosensors, Kuscu2016ModelingReceiver, Bicen2013System-TheoreticCommunication} have been the center of attention due to their controllable environment \cite{R.Aris1956OnTube, Dutta2006EffectSystems}, their ability to simulate biochemical intra-body communications \cite{Femminella2015ADelivery}, and their ability to test lab-on-a-chip technologies \cite{Marx2012Human-on-a-chipMan}. The system is also a suitable platform for testing various modulation, diffusion, and demodulation techniques of IM \cite{Kuscu2019TransmitterTechniques}, making the microfluidic channel ideal for both fundamental research and exploration of potential practical applications. However, the microfluidic MC system still faces a wide variety of challenges that need to be addressed, particularly in detection and noise mitigation \cite{Kuscu2016ModelingReceiver, Llatser2013DetectionCommunication}, before it can be deployed for more practical use cases in the real world.

The stochastic nature of diffusion in MC means that it is also susceptible to many of the same issues that affect traditional electromagnetic (EM) communications, particularly noise interference. This includes different types of noise, such as intersymbol interference (ISI) \cite{Shih2012ChannelCommunications}, environmental noise, binding noise, and Brownian noise \cite{Kuscu2019TransmitterTechniques}, which can lead to high bit error rates (BER) and, as a result, affect the reliability of information transmission \cite{Pierobon2011NoiseNanonetworks}. In addition to the complexity of the microfluidic environment in the MC channel, detection schemes such as maximum likelihood (ML) \cite{Fang2018MaximumCommunication, Kuscu2018MaximumThings} or optimal detection thresholds \cite{Noel2014OptimalNoise} alone may not always provide the most optimal performance under rapidly changing channel conditions and, as demonstrated in \cite{He2016ImprovingAlgorithm, Damrath2016Low-ComplexityCommunication, He2015AdaptiveCommunication, Shrivastava2020AdaptiveCommunication}, adaptive thresholding methods perform promisingly well. Although more recent and complex schemes such as machine learning (ML)-based adaptive thresholding could potentially provide better performance under varying channel conditions \cite{Debus2023ReinforcementMobility, Bartunik2022UsingCommunication, Lee2017MachineCommunications, Yilmaz2017ACommunications, Kim2023ASystem, Qian2018ReceiverNetworks, Qian2019MolecularOptimization}, computational costs, prior knowledge requirements of channel models, and large amounts of learning data requirement may outweigh the benefits in some scenarios \cite{Jing2024PerformanceCommunication}, such as deployment on an intra-body nanomachine. Such a device would ideally require accurate, efficient, and simple detection schemes.

To address these challenges, we propose a novel adaptive thresholding technique based on a Proportional-Integral-Derivative (PID) controller \cite{Wang2020BasicsControl}. In this paper, it will be called an adaptive real-time threshold receiver (ART-Rx). PID controllers are already ubiquitous in various industrial settings and are known to be reliable, fast, and efficient. Unlike ML, it does not require any training or learning of channel models \cite{Wang2020BasicsControl, KiamHeongAng2005PIDTechnology}. Our proposed approach aims to utilize the benefits of PID controllers to adaptively adjust receiver detection thresholds in response to dynamical channel noise conditions. The closed-loop system will allow the receiver to adjust the detection threshold in discrete real-time based on various feedback parameters such as the BER of previously transmitted bits and channel environmental parameters. The implementation of ART-Rx tested in this paper with the Smoldyn simulator \cite{Andrews2012SpatialSimulator} employs a PID controller that adjusts the detection threshold by calculating the error between the observed peak receptor binding levels and a desired setpoint at each symbol interval. The suggested implementation can potentially improve the overall reliability and robustness of the molecular communication receiver in noisy microfluidic environments such as within the human body while remaining computationally efficient.

The remainder of this paper is organized as follows. Section II provides a detailed overview of our ART-Rx implementation and the underlying assumptions. In Section III, we describe the simulation setup, including evaluation parameters and specifications, and present simulation results comparing ART-Rx performance to the statistically optimal detection threshold compared to as a benchmark under various noise levels and channel conditions. This section also includes a comprehensive performance analysis and discusses limitations of our approach, such as the impact of PID parameter tuning and challenges associated with real-world implementation. Finally, Section IV concludes the paper by summarizing key findings and outlines directions for future research on the ART-Rx system.

\section{ART-Rx Implementation}

\begin{figure}[t]
  \centering
  \includegraphics[width=0.4893\textwidth]{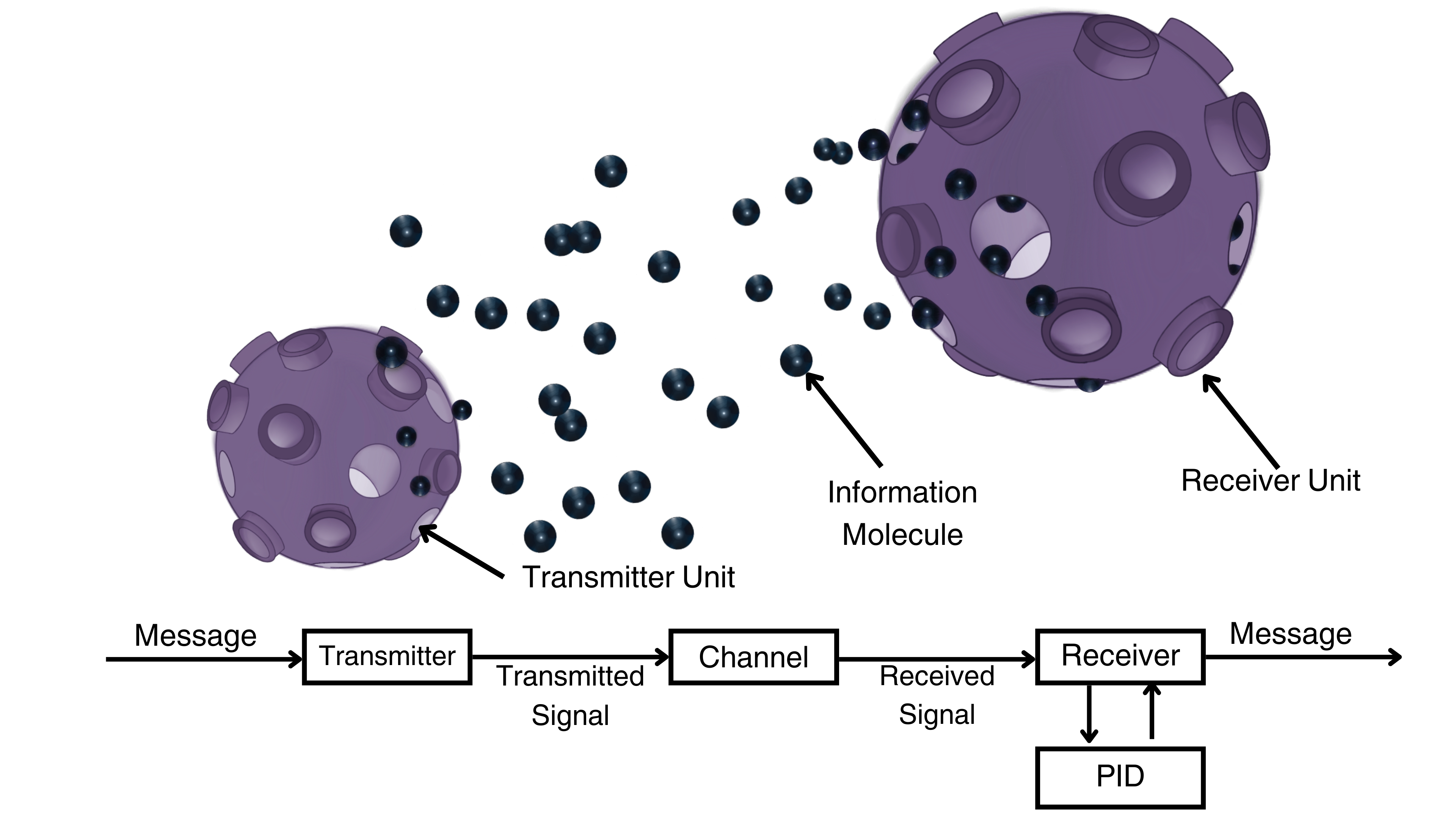}
  \caption{Tx and Rx nanomachines performing MC-based information transfer \cite{Kuscu2016OnBiosensors}.}
  \label{fig:1}
\end{figure}

\begin{figure}[t]
  \centering
  \includegraphics[width=0.4893\textwidth]{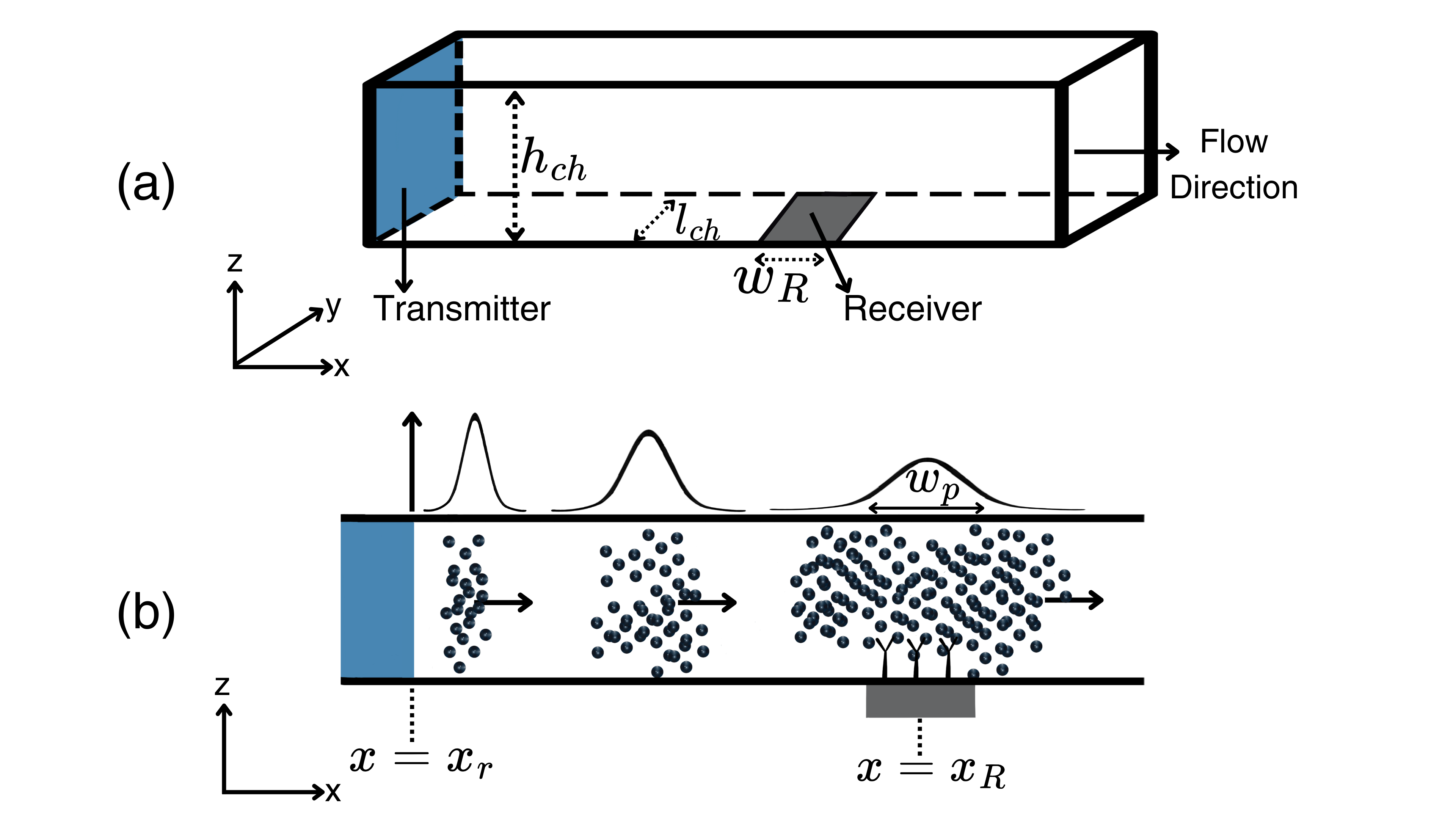}
  \caption{(a) 3-Dimensional and (b) 2-Dimensional view of a rectangular microfluidic propagation channel. The transmitter and receiver locations, together with the dispersion of ligands as they propagate across the channel, are illustrated \cite{Kuscu2016ModelingReceiver}.}
  \label{fig:2}
\end{figure}

The reliable detection of molecular signals in MC receivers is hindered by the stochastic nature of molecular diffusion and variability in microfluidic environments. To address these challenges, we present ART-Rx, an MC receiver incorporating a PID controller with gain scheduling that dynamically adjusts the detection threshold in discrete real-time. The PID controller continuously monitors the observed signal, computes an error signal based on deviation from a desired setpoint, and adjusts the detection threshold to minimize this error. This adaptive mechanism improves system responsiveness, reduces BER, and ensures robust communication under varying channel conditions.

\subsection{Microfluidic MC Setup}

A complete microfluidic MC setup includes a transmitter unit (Tx), a receiver unit (Rx), and a channel through which the IM diffuses to travel from Tx to Rx \cite{Kuscu2016OnBiosensors, Kuscu2016ModelingReceiver, Kuscu2019TransmitterTechniques}. In the literature, nanomachines have been used in place of Tx and Rx, as seen in Figure \ref{fig:1} \cite{Kuscu2016OnBiosensors}. Figure \ref{fig:2} shows a microfluidic MC channel with a rectangular cross-section that was built in Smoldyn and used to test ART-Rx in this paper \cite{Kuscu2016ModelingReceiver}. A Tx which releases IM uniformly across the cross-section resides on the left of the channel. The IM then propagates by convection and diffusion down the channel. An Rx is located further down the channel, at the bottom, populated with BioFET receptors pointing along the z-axis with the purpose of binding to the propagating IM.

\subsection{Modulation Schemes}

There are many well-established modulation schemes \cite{Kuscu2016ModelingReceiver, Kuscu2019TransmitterTechniques, Kuran2011ModulationNanonetworks} for use in MC Tx, including on / off keying (OOK), concentration shift keying (CSK), and molecular shift keying (MoSK). OOK, being one of the simplest modulation schemes, encodes information through the presence or absence of IM during a bit interval \cite{Kuscu2019TransmitterTechniques}. A bit '1' is represented by the release of a predefined number of IM at a bit interval, whereas bit '0' is represented by the absence of IM. CSK uses predefined concentration threshold levels to represent different symbols \cite{Kuscu2019TransmitterTechniques}. As an example, if we set the threshold for bit '1' at 8000 IM and bit '0' at 2000 IM, the respective numbers of IM are released at the beginning of each signal corresponding to the bit being transmitted. MoSK is a modulation scheme that was introduced more recently compared to OOK and CSK \cite{Kuscu2016ModelingReceiver, Kuscu2019TransmitterTechniques}. In MoSK, different types of molecules are used to represent different symbols or bits \cite{Kuscu2019TransmitterTechniques}. Two different types of IM each represent different bits. This allows for improved symbol detection accuracy because the Rx can differentiate between different types of IM, rather than relying on varying molecule concentrations. Through the use of multiple types of molecules to represent more symbols, MoSK can also enable higher data rates \cite{Kuscu2019TransmitterTechniques}. However, this added complexity makes it more challenging to use in certain biological environments. Considering the nature of this, CSK was chosen as the modulation scheme to verify the performance of ART-Rx in this paper.
    
\subsection{Algorithmic Foundations and Working Principles}

\begin{figure}[t]
  \centering
  \includegraphics[width=0.4893\textwidth]{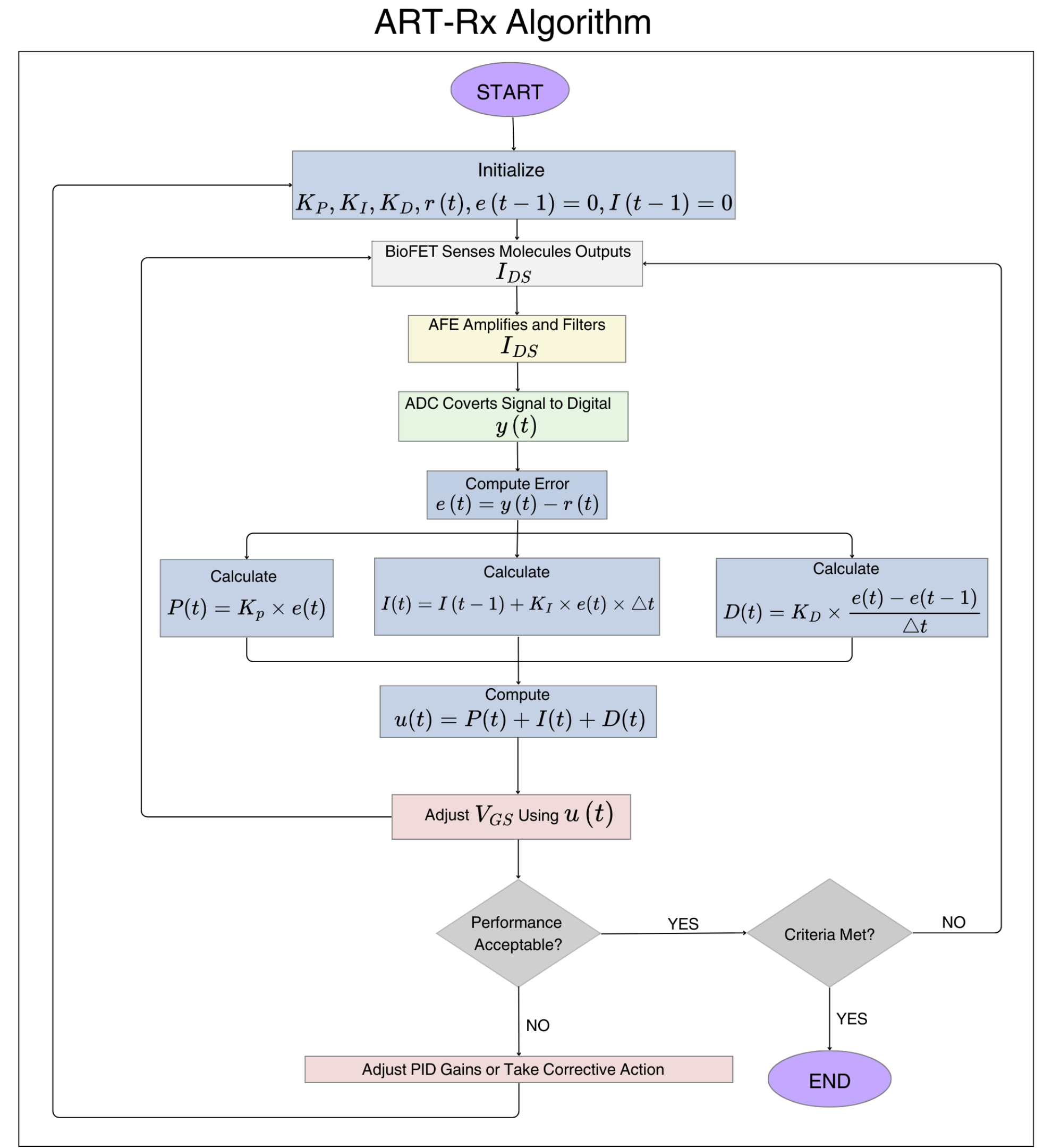}
  \caption{System-on-Chip (SoC) operations algorithm flow diagram with ART-Rx implemented. The diagram illustrates the sequential processes within the SoC, starting with system initialization and proceeding through molecular detection, signal conditioning, analog-to-digital conversion, and digital signal processing.}
  \label{fig:4}
\end{figure}

The algorithm of ART-Rx is based on control theories and consists of three main components: the proportional term, the integral term, and the derivative term. Each of the components contributes to the control signal that ultimately adjusts the Rx detection threshold in a manner that accommodates instantaneous corrections and long-term optimizations of the system. Figure \ref{fig:4} presents a flow diagram of the algorithm behind ART-Rx.

\paragraph{Error Calculation}

At each symbol interval \( t \), the error \( e(t) \) is calculated as the difference between the peak receptor binding level observed \( y(t) \) and the desired setpoint \( r(t) \), i.e.,
\vspace{-4pt}
\begin{equation}
e(t) = y(t) - r(t).
\end{equation}

\noindent The observed value \( y(t) \) represents the maximum number of bound receptors during the symbol interval, which is directly related to the drain-source current \( I_{\text{DS}} \) of BioFET-based Rx \cite{Kuscu2016ModelingReceiver}. The setpoint \( r(t) \) is a reference value that represents the optimal level of receptor binding for accurate detection. This can be determined based on prior knowledge of the system or, in other cases, dynamically updated using statistical measures such as a moving average \cite{Hansun2013AAnalysis}.

\paragraph{Proportional Term}

The proportional term \( P(t) \) provides the system with an immediate corrective action that is proportional to the current error.
\vspace{-4pt}
\begin{equation}
P(t) = K_P \times e(t),
\end{equation}

\noindent where \( K_P \) is the proportional gain. \( P(t) \) makes adjustments to the detection threshold directly proportional to the magnitude of the error \cite{Wang2020BasicsControl, KiamHeongAng2005PIDTechnology}, allowing an initial rapid response of the system to changes in the observed signal. A higher value of \( K_P \) will increase the system's sensitivity to the current error, but if not properly tuned, may lead to an overshoot.

\paragraph{Integral Term}

The role of the integral term \( I(t) \) is to address the accumulated past errors in the system, effectively eliminating offsets and biases in the steady state.
\vspace{-4pt}
\begin{equation}
I(t) = K_I \sum_{k=1}^{t} e(k) \Delta t,
\end{equation}

\noindent where \( K_I \) is the integral gain and \( \Delta t \) is the time interval between symbol transmissions. Through the integration of errors over time, \( I(t) \) corrects for any persistent discrepancies between the setpoint and the observed value, steering the system toward global optimality through the minimization of cumulative errors. However, an excessively high \( K_I \) can lead to negative effects within the system such as integral wind-up and instability \cite{Wang2020BasicsControl, KiamHeongAng2005PIDTechnology}.

\paragraph{Derivative Term}

The role of the derivative term \( D(t) \) is to predict future errors by observing the rate of change in errors within the system.
\vspace{-4pt}
\begin{equation}
D(t) = K_D \left( \frac{e(t) - e(t-1)}{\Delta t} \right),
\end{equation}

\noindent where \( K_D \) is the derivative gain. \( D(t) \) enhances the system's responsiveness by providing a damping effect which reduces overshoot and prevents oscillations. The derivative term anticipates future errors based on the current error's rate of change, contributing to system stability and achieving local optimality by minimizing instantaneous deviations \cite{Wang2020BasicsControl, KiamHeongAng2005PIDTechnology}.

\paragraph{Control Signal and Threshold Adjustment}

We can therefore obtain the control signal \( u(t) \) from the sum of the proportional, integral, and derivative terms, i.e.,
\vspace{-4pt}
\begin{equation}
\begin{split}
u(t) = P(t) + I(t) + D(t) &= K_P \times e(t) + K_I \sum_{k=1}^{t} e(k) \Delta t \\
&\quad + K_D \left( \frac{e(t) - e(t-1)}{\Delta t} \right).
\end{split}
\end{equation}

\noindent The detection threshold \( \theta(t) \) is updated iteratively according to the control signal, i.e.,
\vspace{-4pt}
\begin{equation}
\theta(t+1) = \theta(t) + u(t).
\end{equation}

\noindent The updated threshold is then used for the detection of the subsequent symbol, enabling the system to adapt to changes in the MC channel in discrete real-time.

\paragraph{Avoiding Integral Windup}

To avoid integral windup, which happens when the integral term accumulates excessively due to sustained errors, anti-windup strategies are implemented by clamping the integral term within predefined limits, i.e.,
\vspace{-4pt}
\begin{equation}
I(t) = \mathrm{clamp}\left( I(t-1) + e(t) \Delta t, I_{\text{min}}, I_{\text{max}} \right),
\end{equation}

\noindent where \( \mathrm{clamp}(x, a, b) \) restricts \( x \) within the ranges of \([a, b]\), and \( I_{\text{min}} \), \( I_{\text{max}} \) are the minimum and maximum allowed integral values, respectively.

\subsection{Local and Global Optimality}

The ART-Rx system aims to achieve both local and global optimality in threshold adjustments made through the PID controller.

\paragraph{Local Optimality}

Local optimality involves optimizing the performance of the system at each individual time step, primarily influenced by the proportional \( P(t) \) and derivative \( D(t) \) terms of the PID controller \cite{KiamHeongAng2005PIDTechnology}. By minimizing instantaneous error \( e(t) \) and damping error fluctuations, the controller quickly adjusts the detection threshold in response to rapid changes in the signal. The derivative term \( D(t) \) is especially crucial, as it anticipates future errors based on current system trends \cite{KiamHeongAng2005PIDTechnology}.

\paragraph{Global Optimality}

Global optimality focuses on optimizing system performance over an extended period, minimizing cumulative error, and overall BER. The integral term \( I(t) \) plays a significant role in the integration of the error over time, correcting persistent biases and drifts \cite{KiamHeongAng2005PIDTechnology}. This integration ensures that the detection threshold converges to a value that minimizes the average BER throughout the transmission.

Balancing the contributions of the proportional, integral, and derivative terms is essential. Proper tuning of PID gains \( K_P \), \( K_I \), and \( K_D \), along with techniques such as gain scheduling to address nonlinearities \cite{Astrom1993AutomaticSurvey}, ensures that the controller effectively manages immediate and long-term discrepancies, maintains system stability, and optimizes detection performance.

\subsection{SoC Architecture and Physical Implementation}

\begin{figure*}[t]
  \centering
  \includegraphics[width=1\textwidth]{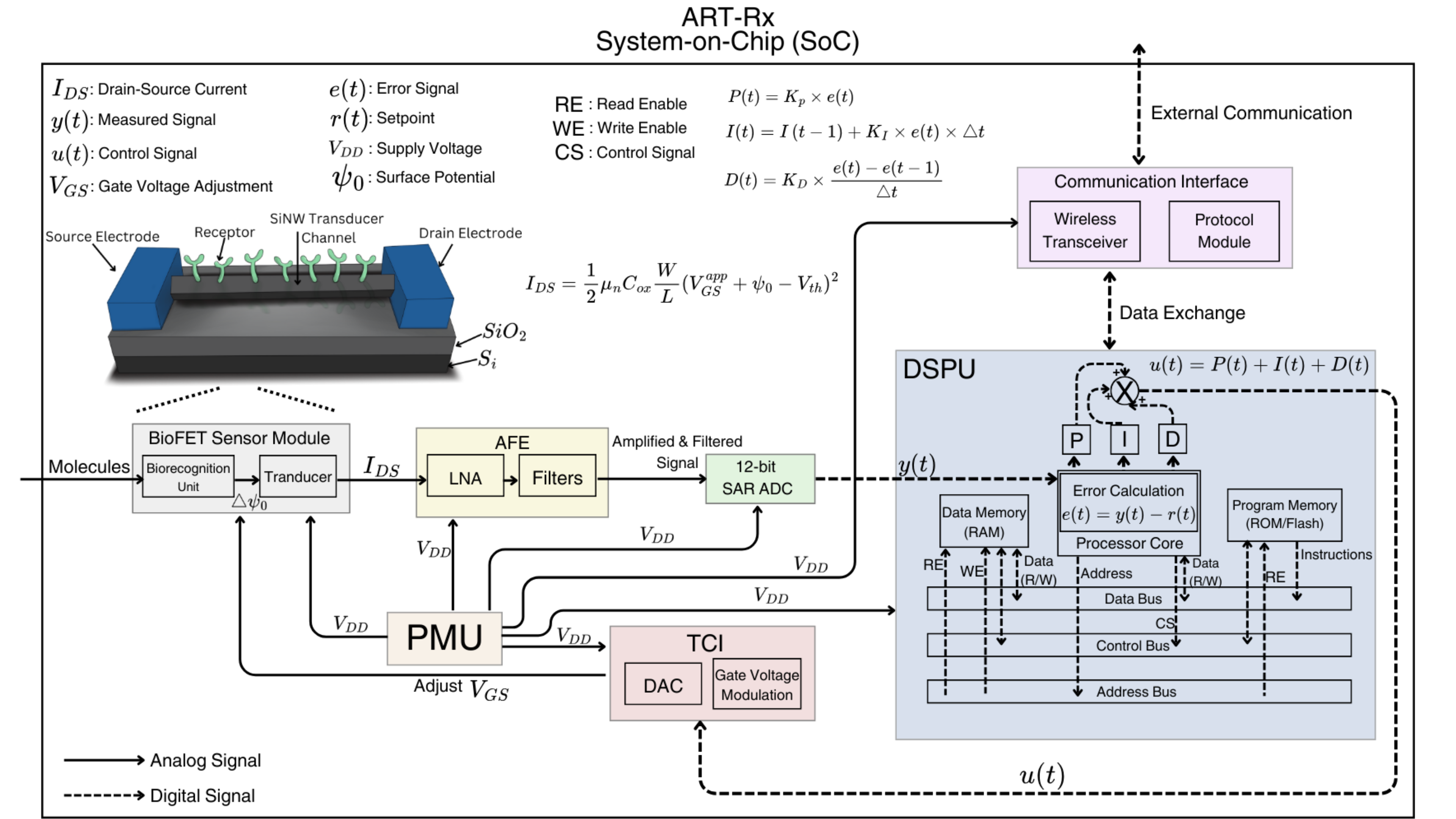}
  \caption{The figure depicts the integrated components within the SoC boundary: the BioFET Sensor Module with its Biorecognition Unit and Transducer \cite{Kuscu2016ModelingReceiver}; the Analog Front-End (AFE) comprising the Low Noise Amplifier (LNA) and Filters; and the Analog-to-Digital Converter (ADC). The Digital Signal Processing Unit (DSPU) is detailed with its Processor Core, Program Memory (ROM/Flash) storing the code, and Data Memory (RAM) storing parameters such as the setpoint $r(t)$, PID gains $K_P$, $K_I$, $K_D$, and previous values. Data transfer and control signals are managed via the Data Bus, Address Bus, and Control Bus, connecting the Processor Core with the Memory components. The DSPU computes the error signal $e(t)$ and processes it through the PID controller to generate the control signal $u(t)$, which is sent to the Threshold Control Interface (TCI). This interface adjusts the gate voltage $V_{\text{GS}}$ of the BioFET sensor, forming a feedback loop that enhances system responsiveness and accuracy. The Communication Interface enables data exchange between the SoC and external systems.}
  \label{fig:5}
\end{figure*}

To physically realize the ART-Rx system, we propose a system-on-chip (SoC) architecture that integrates all essential components onto a single chip. This integration is crucial at the nanoscale because of space and power constraints and the need for efficient signal processing. Figure~\ref{fig:5} shows a high-level SoC design of an MC Rx with the proposed ART-Rx implemented.

\paragraph{BioFET Sensing Module}

The BioFET sensing module serves as the core sensor component, acting as both the biorecognition unit and the transducer. A close-up diagram of a typical BioFET is also shown in the figure \cite{Kuscu2016ModelingReceiver}. The gate surface is functionalized with aptamer receptors that selectively bind to specific IM, inducing changes in surface potential \( \psi_0 \) \cite{Kuscu2016OnBiosensors, Kuscu2016ModelingReceiver}. 2-D and 3-D materials such as silicon nanowires or graphene are chosen for the channel due to their high electron mobility, biocompatibility, and compatibility with CMOS fabrication processes \cite{Kuscu2016OnBiosensors}.

The drain-source current \( I_{\text{DS}} \) is given by \cite{Cheng2002MosfetGuide}
\vspace{-4pt}
\begin{equation}
I_{\text{DS}} = \frac{1}{2} \mu_n C_{\text{ox}} \frac{W}{L} \left( V_{\text{GS}}^{\text{app}} + \psi_0 - V_{\text{th}} \right)^2,
\end{equation}

\noindent where \( \mu_n \) is the mobility of the electron, \( C_{\text{ox}} \) is the capacitance of the gate oxide per unit area, \( W \) and \( L \) are the width and length of the channel, respectively, \( V_{\text{GS}}^{\text{app}} \) is the applied gate source voltage and \( V_{\text{th}} \) is the threshold voltage.

\paragraph{Analog Front-End (AFE) and Analog-to-Digital Converter (ADC)}

The AFE amplifies and conditions the BioFET's electrical signals, incorporating low-noise amplifiers (LNA) and filters to enhance signal quality while minimizing noise. The design of the AFE places emphasis on ultra-low power consumption and minimal area usage, which is essential for nanoscale integration of an SoC.

The conditioned analog signals from the AFE are digitized by an ADC. The Successive Approximation Register (SAR) ADC is selected for the SoC because of its favorable and balanced trade-offs between power efficiency, conversion speed, and resolution. A 12-bit resolution is chosen to ensure that the subtle variations in the BioFET's output are all captured, ensuring accurate digitization of the signal for processing by the PID controller further down the pipeline.

\paragraph{Digital Signal Processing Unit (DSPU) with ART-Rx PID Controller Module}

The DSPU is a custom-designed nanoscale processor tailored for low-power operations and fixed-point arithmetic with the aim of reducing computational complexity. It includes memory units to store variables and program data and contributes to the efficient execution of the PID algorithm behind the ART-Rx module. The selection of a digital implementation over an analog implementation offers benefits such as flexibility, programmability, and precise control, which are advantageous for the nanoscale system.

\paragraph{Threshold Control Interface (TCI)}

The TCI adjusts the BioFET detection threshold based on the control signal \( u(t) \) passed on from the PID controller. This is achieved through modulation of the gate voltage \( V_{\text{GS}} \) using a Digital-to-Analog Converter (DAC) together with voltage control circuits. A high-resolution DAC enables precise threshold adjustments, which is critical to maintaining the detection accuracy of the system. The interface is designed to prevent interference with the BioFET's sensing function, with careful layout and isolation techniques employed to avoid crosstalk and EM interference from neighboring components and circuitry.

\paragraph{Power Management Unit (PMU)}

The PMU is designed to efficiently manage the power distribution within the SoC, addressing the efficiency and strict thermal requirements of nanoscale devices. The PMU incorporates nanoscale voltage regulators and utilizes power gating techniques to minimize power consumption across all components within the SoC. By providing stable and adjustable voltage levels, the PMU improves the energy efficiency and reliability of the entire SoC. The integration of the PMU allows for dynamic power management, enabling the SoC to adapt its power usage according to system needs, which is crucial to extend the operational lifespan of the system in resource-constrained environments.

\paragraph{Communication Interface}

The Communication Interface facilitates seamless data exchange between the SoC and external systems or networks. It integrates ultra-low power transceiver circuits designed to support bidirectional communication, enabling features such as remote monitoring, program updates, and data transmission. The modular and programmable design of the Communication Interface provides flexibility to accommodate various communication protocols, enhancing the adaptability and integration capabilities of the SoC. Including this interface ensures that the system can participate effectively in networked environments commonly seen in IoBNT applications, which is advantageous for applications requiring distributed sensing and control.

\paragraph{Monolithic Integration}

Integration of all components in a single package minimizes parasitic capacitance and inductance, improves signal integrity, and reduces interconnect lengths. It also brings other benefits, such as simplification of the fabrication process and improved SoC reliability through the minimization of the number of discrete connections and components.

\paragraph{Material Compatibility and Fabrication Processes}

Using silicon or graphene, which are well-known materials that are compatible with industry standard CMOS fabrication processes, helps facilitate the monolithic integration of the BioFET recognition unit with the AFE, ADC, DSPU, and PID controllers of the ART-Rx module. Compatibility allows the use of established manufacturing techniques, resulting in a reduction in fabrication complexities and costs.

\paragraph{Low-Power Design}

Power efficiency is not only important but also critical for nanoscale devices. Components at the nanoscale are designed for ultra-low power operations using techniques such as power gating, dynamic voltage and frequency scaling (DVFS), and optimized circuit topologies. Enabling the DSPU to use fixed-point arithmetic should reduce computational costs, resulting in lower power usage by the component.

\paragraph{Scalability and Flexibility}

Employing a SoC architecture design enables support for scalability, allowing for the addition of additional functionalities or modifications for adaptation to specific application needs. Furthermore, selecting a digital implementation for the PID controller and the DSPU component allows for flexibility should there be a need to update control algorithms and parameters post-deployment.

\paragraph{Thermal Management and Signal Integrity}

Careful thermal management strategies are implemented to dissipate the heat produced by the SoC components, preventing performance degradation due to variations in temperature. Signal integrity is maintained through shielding structures, differential signaling, and proper grounding techniques.

\subsection{Statistical Optimal Detection Threshold as a Benchmark}

The optimal detection threshold $\lambda_{TH,optimal}$ for our CSK modulation scheme is compared with to evaluate the performance of our ART-Rx module and is calculated based on the statistical properties of bound receptors for bits '0' and '1'. The process involves several steps derived from \cite{Kuscu2016ModelingReceiver}.

1) First, we calculate the molecular concentration $c_m$ for each bit
\vspace{-4pt}
\begin{equation}
c_m = \frac{N_m}{A_{ch} \sqrt{4\pi D t_d}},
\end{equation}

\noindent where $N_m$ is the number of molecules released, $A_{ch}$ is the cross-sectional area of the channel, $D$ is the diffusion coefficient and $t_d$ is the arrival time of the peak concentration.

2) Next, we determine the probability of a receptor bound state in the absence of interference
\vspace{-4pt}
\begin{equation}
P_{B,exp} = \frac{c_m / K_{D,m}}{1 + c_m / K_{D,m}},
\end{equation}

\noindent where $K_{D,m}$ is the dissociation constant of the ligands.

3) We then calculate the mean and variance of the bound receptors for each bit, assuming Poisson statistics
\vspace{-4pt}
\begin{equation}
\begin{split}
\delta I_{mean,i} &= N_r P_{B,exp,i} \\
\delta I_{var,i} &= N_r P_{B,exp,i} (1 - P_{B,exp,i}),
\end{split}
\end{equation}

\noindent where $i \in \{0,1\}$ represents the bit value and $N_r$ is the number of receptors.

4) Finally, we calculate the optimal detection threshold using these statistical properties
\vspace{-4pt}
\begin{equation}
\begin{split}
\lambda_{TH,optimal} = & \frac{1}{\sigma_1^2 - \sigma_0^2} \Big[ (\sigma_1^2 \mu_0 - \sigma_0^2 \mu_1) + \\
& \sigma_1 \sigma_0 \sqrt{(\mu_1 - \mu_0)^2 + 2(\sigma_1^2 - \sigma_0^2) \ln(\frac{\sigma_1}{\sigma_0})} \Big]
\end{split},
\end{equation}

\noindent where $\sigma_i^2 = \delta I_{var,i}$ and $\mu_i = \delta I_{mean,i}$ for $i \in \{0,1\}$.

This optimal threshold $\lambda_{TH,optimal}$ minimizes the probability of error in the CSK modulation scheme by considering the statistical distributions of the bound receptors for each symbol's bit. It also provides a benchmark against which we can compare the performance of our ART-Rx implementation. It will be referred to as the "optimal method" throughout the remainder of this paper.

The PID gains were initially tuned using the Ziegler-Nichols ultimate gain method \cite{Ziegler1993OptimumControllers} in conjunction with empirical methods, where $K_i$ and $K_d$ are set to zero and $K_p$ is increased until sustained oscillations occur at the ultimate gain $K_u$. The final PID gains were then calculated using the measured ultimate gain and oscillation period. Then, fine-tuning was performed through iterative simulations to optimize system response based on a plot of the detection threshold fluctuations. For scenarios with stable channel environmental conditions but varying interferer noise molecules, the final gain values were established at $K_p$ = 0.18, $K_i$ = 0.02, and $K_d$ = 0.005. For other scenarios with varying channel environmental conditions, gain scheduling was used, and individual sets of PID gains were tuned using the Ziegler-Nichols method for each simulation scenario. This tuning process aimed to balance the rapid response to changing channel conditions with system stability and to account for any nonlinearities. To prevent integral windup, we implemented an anti-windup mechanism that clamps the integral term within a predefined limit of 100.

\subsection{Assumptions}

In implementing our PID-controlled ART-Rx system simulated using Smoldyn \cite{Andrews2012SpatialSimulator}, we make several key assumptions.

\paragraph{Negligible External Disturbances}

We assume external disturbances, unmodeled dynamics, and higher-order nonlinearities are negligible in our simulation environment. Smoldyn allows us to minimize these factors, ensuring that the PID controller's performance is not compromised by unforeseen behaviors.

\paragraph{Effective PID Gain Tuning and Gain Scheduling} The PID controller gains are appropriately tuned for different operating regions through gain scheduling to handle the nonlinear effects and ensure satisfactory performance under varying channel conditions. We use both Ziegler-Nichols tuning \cite{Ziegler1993OptimumControllers} and simulation-based optimization to determine optimal gains $K_P$, $K_I$, and $K_D$ for each operating region, minimizing the control error \cite{Astrom1993AutomaticSurvey}, i.e.,
\vspace{-4pt}
\begin{equation}
J = \int_0^T e^2(t) \, dt,
\end{equation}
where $J$ is the performance index and $e(t)$ is the control error.

\paragraph{Linearization of BioFET Sensor Response}

We assume that the nonlinear response of the BioFET sensor can be linearized around operating points. The drain current $I_{\text{DS}}$ is a nonlinear function \cite{Cheng2002MosfetGuide}, i.e.,
\vspace{-4pt}
\begin{equation}
I_{\text{DS}} = \frac{1}{2} \mu_n C_{\text{ox}} \frac{W}{L} \left( V_{\text{GS}}^{\text{app}} + \psi_0 - V_{\text{th}} \right)^2.
\end{equation}
Linearizing around $\psi_0 = \psi_{0,0}$ yields
\vspace{-4pt}
\begin{equation}
\Delta I_{\text{DS}} = G_m \Delta \psi_0,
\end{equation}
where $G_m$ is the transconductance at the operating point
\vspace{-4pt}
\begin{equation}
G_m = \mu_n C_{\text{ox}} \frac{W}{L} \left( V_{\text{GS}}^{\text{app}} + \psi_{0,0} - V_{\text{th}} \right).
\end{equation}
This piecewise-linear relationship allows us to use the sensor output in our gain scheduling control framework.

\section{Performance Analysis}

To fully evaluate the performance of the MC system with our ART-Rx implemented, simulations were performed using MATLAB and Smoldyn \cite{Andrews2012SpatialSimulator}. The data were mainly analyzed using bit error rate (BER) and bit error probability (BEP), which are reliable indicators of the system's accuracy in decoding symbols. To benchmark and compare the performance of our ART-Rx system, simulations were run simultaneously using the optimal detection threshold method, as introduced in Section II and mathematically defined by (12). Referred to as the "optimal method", this method, which calculates a static threshold that minimizes the probability of error based on the statistical properties of the received signal, serves as a standard benchmark here because of its theoretical optimality under static channel conditions. Figure \ref{fig:2} shows the microfluidic MC channel that was built in Smoldyn to run the simulations. Table \ref{tab:I} shows the values of the parameters configured for the simulation, based on those used in \cite{Kuscu2016ModelingReceiver}. Noise is represented in the simulation as interferer molecules (numI) that also diffuse together with the IM undergoing Brownian motion and may incorrectly attach onto the receptors. The detection threshold is updated for every transmitted bit.

\begin{table}[h]
\centering
\caption{Simulation Parameters}
\label{tab:I}
\begin{tabular}{|l|l|l|}
\hline
\textbf{Parameter} & \textbf{Symbol} & \textbf{Value} \\
\hline
\textit{Microfluidic channel height}& $h_{ch}$& 5 $\mu$m \\
\hline
\textit{Microfluidic channel width} & $w_{ch}$ & 10 $\mu$m \\
\hline
\textit{Microfluidic channel length} & $l_{ch}$ & 200 $\mu$m \\
\hline
\textit{Average flow velocity} & $u$ & 10 $\mu$m/s \\
\hline
\textit{Receiver's center position} & $x_r$ & 3 mm \\
\hline
\textit{Intrinsic diffusion coefficient} & $D_0$ & $20 \, \text{pm}^2/\text{s}$ \\
\hline
\textit{Binding rate of ligands} & $K_{b_m}$ & $2 \times 10^{-17} {m}^3/s$ \\
\hline
\textit{Unbinding rate of ligands} & $K_{u_m}$ & 1 $\text{/s}$ \\
\hline
\textit{Number of independent receptors} & $N_r$ & 200 \\
\hline
\textit{Length of the graphene channel} & $l_{gr}$ & 5 $\mu$m\\
\hline
\textit{Width of the graphene channel} & $w_{gr}$ & 10 $\mu$m \\
\hline
\textit{PID: Proportional} & $K_p$ & 0.18 \\
\hline
\textit{PID: Integral} & $K_i$ & 0.02 \\
\hline
\textit{PID: Derivative} & $K_d$ & 0.005 \\
\hline
\textit{Symbols transmitted} & $numSymbol$ & 100 \\
\hline
\end{tabular}
\end{table}

\subsection{Implementation in Simulation Code}

The key components of the simulation include:

\begin{itemize}
    \item \textit{Bitstream Generation and Modulation}: A random bitstream is generated, and each bit is modulated using CSK, where different concentrations of IM represent binary symbols of '0' or '1'.
    \item \textit{MC Channel Simulation}: The emission, diffusion, and reception of signaling molecules are simulated, taking into account factors such as diffusion coefficients, flow velocity, and binding kinetics characterized by the binding rate \( k_{\text{bind}} \) and the non-binding rate \( k_{\text{unbind}} \).
    \item \textit{BioFET Sensor Modeling}: The BioFET response to molecular concentration is modeled, incorporating the relationship between surface potential \( \psi_0 \), gate source voltage \( V_{\text{GS}} \), and drain source current \( I_{\text{DS}} \).
\end{itemize}

At each symbol interval, the simulation records the maximum number of bound receptors \( y(t) \). The PID controller computes the error \( e(t) \), calculates the control signal \( u(t) \), and updates the detection threshold \( \theta(t+1) \). The updated threshold is then used to detect the transmitted bit by comparing it with \( y(t) \). The setpoint \( r(t) \) is dynamically updated based on statistical measures and predefined optimal values.

\paragraph{PID Gain Tuning}

The PID gains are tuned through simulation-based optimization. The Ziegler-Nichols tuning method \cite{Ziegler1993OptimumControllers} is used together with iterative adjustments to achieve the desired system performance. For scenarios with stable channel environmental conditions but varying interferer noise molecules, the final gain values were established at $K_p$ = 0.18, $K_i$ = 0.02, and $K_d$ = 0.005. The gains are selected to ensure system stability, fast response, minimal overshoot, and low steady-state error. For other scenarios with varying channel environmental conditions, gain scheduling was used, and individual sets of PID gains were tuned using the Ziegler-Nichols method for each simulation scenario. This tuning process aimed to balance the rapid response to changing channel conditions with system stability and to account for any nonlinearities. To prevent integral windup, we implemented an anti-windup mechanism that clamps the integral term within a predefined limit of 100.

\paragraph{Initial System Threshold}

Starting with an initial threshold $\gamma_0$ in the system, which is set based on the expected probabilities of receptor binding and noise characteristics.
\vspace{-4pt}
\begin{equation}
\gamma_0 = N_r \left(\frac{P_{B,\text{exp},0} + P_{B,\text{exp},1}}{2} + \frac{\mu_I}{\sigma_I^2}\right),
\end{equation}

\noindent where $N_r$ is the number of independent receptors, $P_{B,\text{exp},0}$ and $P_{B,\text{exp},1}$ are the expected probabilities of the bound state of a receptor for bits '0' and '1,' respectively, $\mu_I$ is the mean noise and $\sigma_I^2$ is the variance of the noise.

This initial threshold takes into account both the expected receptor binding states for bits '0' and '1' together with the characteristics of the noise in the system. This provides a starting point from which our PID controller can then adapt to optimize performance.

\paragraph{BER Calculation Method}

The calculation of the current BER, $\text{BER}_c(i)$ is computed using a bit-by-bit comparison for error detection, i.e.,
\vspace{-4pt}
\begin{equation}
\text{BER}_c(i) = \frac{1}{M} \sum_{i=1}^M |b_i - \hat{b}_i|,
\end{equation}

\noindent where $b_i$ is the $i$-th transmitted bit and $\hat{b}_i$ is the $i$-th decoded bit.

\paragraph{Advantages of Observed Value-Based Error Calculation}

The observed value-based error \( e(t) = y(t) - r(t) \) can be obtained and used to provide immediate feedback at each symbol interval, allowing the PID controller to adjust the detection threshold in discrete real time. This approach is computationally efficient and aligns with the principles of control theory \cite{Wang2020BasicsControl, KiamHeongAng2005PIDTechnology}. Although BER could also be used as feedback error for the PID controller, it is thought that the error signal introduces latency due to the need to accumulate sufficient data to compute BER, reducing system responsiveness and increasing computational costs, which can be difficult for devices of nanoscale with limited resources.

\begin{figure*}[t]
    \centering
    \subfloat[]{
    \includegraphics[width=0.32\textwidth]{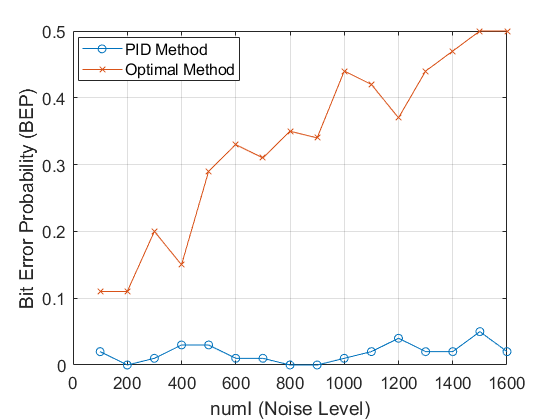}
}
\subfloat[]{
    \includegraphics[width=0.32\textwidth]{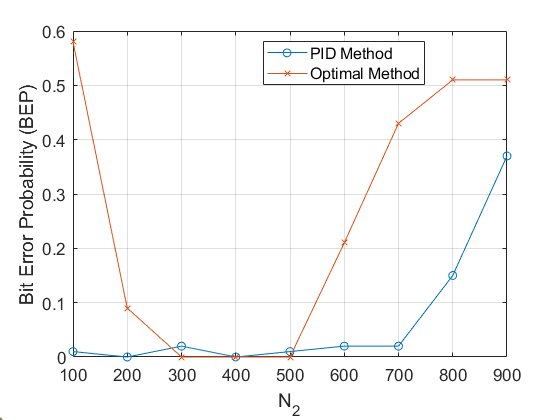}
}
\subfloat[]{
    \includegraphics[width=0.32\textwidth]{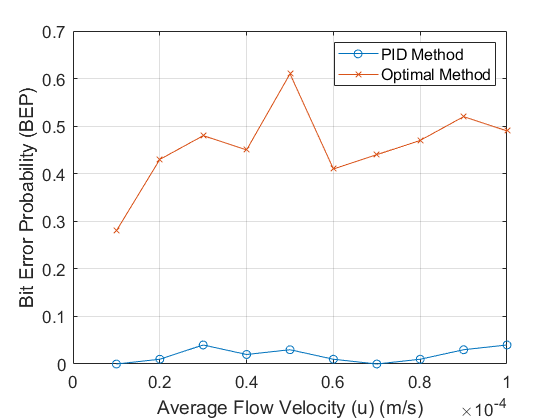}
}
\\
\subfloat[]{
    \includegraphics[width=0.32\textwidth]{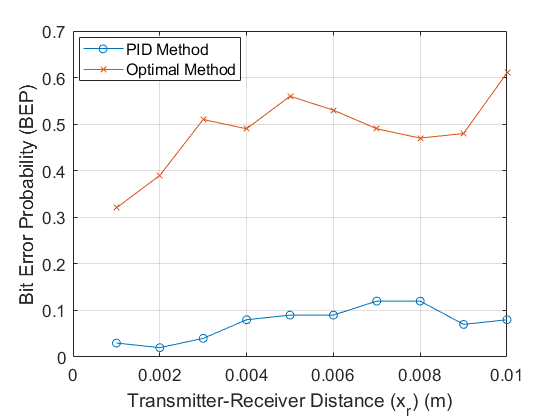}
}
\subfloat[]{
    \includegraphics[width=0.32\textwidth]{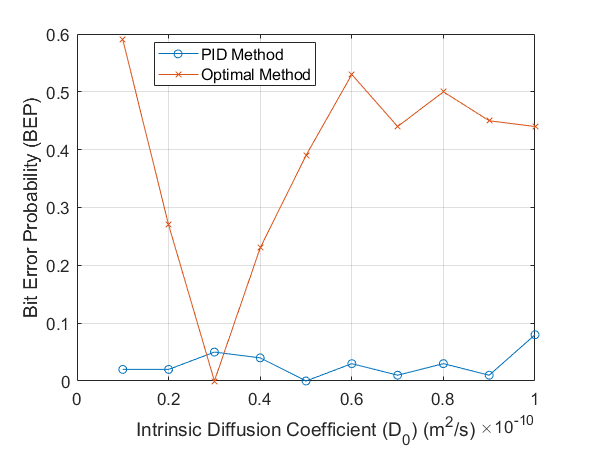}
}
\subfloat[]{
    \includegraphics[width=0.32\textwidth]{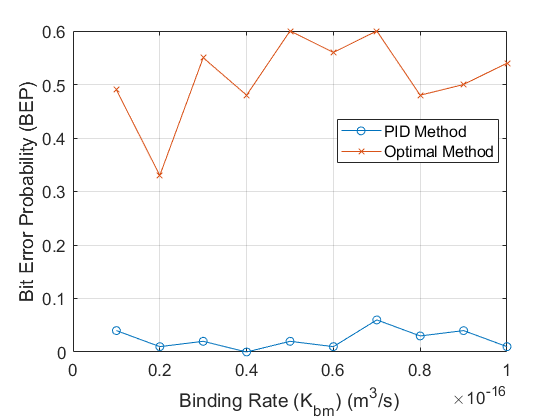}
}
    \caption{BEP performance comparison between PID-based ART-Rx and the optimal method for varying (a) interferer molecule count, (b) $N_2$ CSK concentration levels, (c) average flow velocity, (d) transmitter-receiver distance, (e) intrinsic diffusion coefficient, (f) binding rate. In Figures (b)-(d), the concentration levels of CSK modulation were set at $N_1$ = 1000 for bit '1' and $N_2$ = 600 for bit '0' with the interferer molecule count numI set at 700. For Figure (a), the same concentration level settings were used but with varying interferer molecule count numI from 100 to 1600.}
    \label{fig:6}
\end{figure*}

\begin{figure*}[t]
    \centering
    \subfloat[]{
    \includegraphics[width=0.32\textwidth]{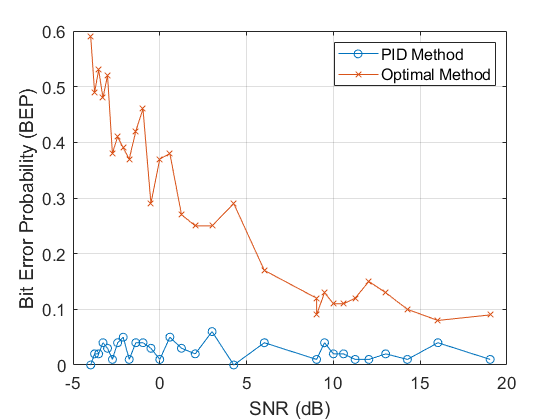}
}
\subfloat[]{
    \includegraphics[width=0.32\textwidth]{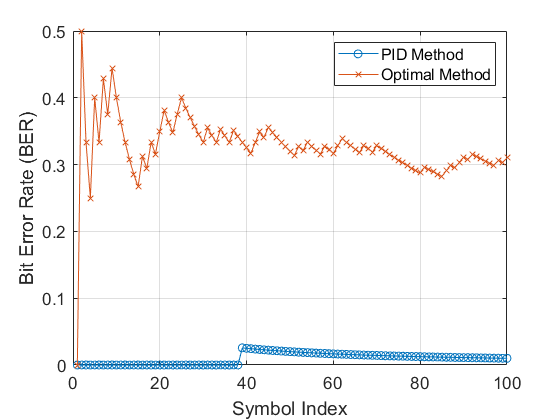}
}
\subfloat[]{
    \includegraphics[width=0.32\textwidth]{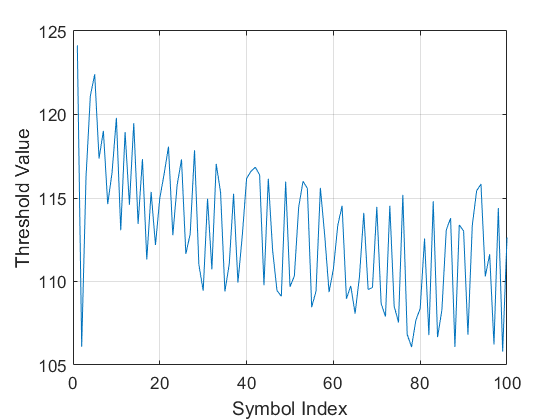}
}
    \caption{Performance comparison between PID-based ART-Rx and the optimal method of (a) BEP against SNR, (b) BER against 100 symbol indexes when the interferer molecule count numI is set at 700, (c) detection threshold adjustment logs of the PID controller when the interferer molecule count numI is set at 700. For Figures (a)-(c), the CSK concentration levels were established at $N_1$ = 1000 for bit '1' and $N_2$ = 600 for bit '0'.}
    \label{fig:7}
\end{figure*}

\subsection{BEP vs. Varying Interferer Molecules}

Using fixed microfluidic channel parameters from Table \ref{tab:I}, Figure \ref{fig:6}(a) presents a comprehensive comparison of BEP performance between our PID-based ART-Rx and the optimal method. The evaluation is carried out on varying numbers of interferer molecules, ranging from 100 to 1600. It can be observed that the BEP of the optimal method gradually increases from 0.11 at numI = 100 to 0.5 at numI = 1600, which is an expected result considering the proportionate increase of interferer molecules, which increases the sensor module's difficulty in accurately decoding the symbol. In contrast, our ART-Rx had BEPs of no more than 0.05 in all numI ranges, significantly outperforming the optimal method. The overall average BEP for these results was 0.33 for the optimal method and 0.018 for ART-Rx.

Figure \ref{fig:7} further presents results related to this section. Figure \ref{fig:7}(a) shows the BEP performance at various levels of SNR ranging from -5 to 20. A downward BEP trend from 0.59 to 0.09 can be clearly observed for the optimal method as the SNR increases. In contrast, our ART-Rx implementation performed exceptionally well with very low BEP values of no higher than 0.06 across all SNR levels. To provide more insight on the inner workings of the PID controller, Figure \ref{fig:7}(b) presents a BER performance graph for a numI value of 700 in all 100 transmitted symbols. Once again, the performance gap between the optimal method and our ART-Rx is immediately noticeable with incorrectly decoded bits throughout the transmission duration of the optimal method with a final BER of 0.31. In contrast, our ART-Rx only had one incorrectly decoded bit throughout the entire transmission duration at symbol index 39 with a final BER of 0.01. Figure \ref{fig:7}(c) presents an overview of the adjustments made to the detection threshold values based on the output of the PID controller. In all of our findings, it was observed that scenarios with lower interferer molecule counts resulted in a higher detection threshold, and scenarios with higher interferer molecule counts had a lower detection threshold. This is to be expected because the more interferer molecules there are in the channel, the more incorrect binding events there are, and hence, less IM gets to correctly bind to its dedicated receptors, and the PID controller compensates for it. This behavior also verifies that the system is working properly. For Figure \ref{fig:7}(c) specifically where numI was set at 700, the detection threshold oscillates between 105 and 122.

In general, the robustness of our PID method is particularly evident in challenging conditions where there is a larger number of interferer molecules. The results here demonstrate the ability and effectiveness of our PID-based ART-Rx approach in adapting to varying levels of interference molecules in the MC microfluidic channel.

\subsection{BEP vs. Varying CSK Concentration Levels}
In this section, we further evaluated the BEP performance of our ART-Rx by setting a fixed $N_1$ concentration at 1000 for the CSK modulation scheme and varying $N_2$ concentrations between 100 and 900. By conducting this test, we can observe how our system performs with CSK when the predefined concentration levels are far apart and close together. As seen in Figure \ref{fig:6}(b), although our PID method started to struggle at 800 and beyond, overall it still maintained a consistently low BEP while iterating through all values of $N_2$. The optimal method also performed exceptionally well for values of $N_2$ between 300 and 500. Our findings suggest that ART-Rx is also flexible in operating with a wider range of predefined concentration levels for CSK modulation, potentially opening up the possibilities of having more predefined concentration levels for 2-bit transmissions, significantly increasing throughput and bandwidth. However, as the results suggest, it seems that our current PID configuration performs best when CSK concentration levels $N_1$ and $N_2$ are defined at least 300 or 30\% apart. The overall average BEP was 0.259 for the optimal method and 0.067 for the PID method.

\subsection{BEP vs. Varying Average Flow Velocity}
This section continues to analyze the BEP performance of our ART-Rx with varying environmental conditions within the microfluidic channel, starting with the varying average flow velocities ($u$) presented in Figure \ref{fig:6}(c). As flow velocity directly influences channel dynamics and IM movement, Smoldyn simulations were performed for varying mean flow velocities from $1 \times 10^{-5}$ m/s to $10 \times 10^{-5}$ m/s to investigate how it would affect our ART-Rx implementation. As the channel dynamics changed with the variations of the channel parameters, gain scheduling was implemented for the PID controller to better conform to the nonlinear process. Different gains $K_p$, $K_i$, and $K_d$, obtained using Ziegler-Nichols and empirical tuning methods, were used and are listed in Table \ref{tab:II}. Fixed CSK concentration levels of $N_1$ = 1000 and $N_2$ = 600 were used together with numI = 700. The higher interferer molecule count of 700 was selected to evaluate the system's performance simultaneously in varying channel environmental conditions and high-noise interference conditions.

\begin{table}[h]
\centering
\caption{PID Gain Parameters for Average Flow Velocity Simulations}
\label{tab:II}
\begin{tabular}{|l|l|}
\hline
\textbf{Average Flow Velocity} & \textbf{PID Parameters} \\
\hline
$1 \times 10^{-5}$ m/s & $K_p$ = 0.18, $K_i$ = 0.02, $K_d$ = 0.005 \\
\hline
$2 \times 10^{-5}$ m/s & $K_p$ = 0.15, $K_i$ = 0.02, $K_d$ = 0.01 \\
\hline
$3 \times 10^{-5}$ m/s & $K_p$ = 0.15, $K_i$ = 0.02, $K_d$ = 0.01 \\
\hline
$4 \times 10^{-5}$ m/s & $K_p$ = 0.15, $K_i$ = 0.02, $K_d$ = 0.01 \\
\hline
$5 \times 10^{-5}$ m/s & $K_p$ = 0.13, $K_i$ = 0.025, $K_d$ = 0.015 \\
\hline
$6 \times 10^{-5}$ m/s & $K_p$ = 0.13, $K_i$ = 0.025, $K_d$ = 0.015 \\
\hline
$7 \times 10^{-5}$ m/s & $K_p$ = 0.13, $K_i$ = 0.025, $K_d$ = 0.015 \\
\hline
$8 \times 10^{-5}$ m/s & $K_p$ = 0.12, $K_i$ = 0.03, $K_d$ = 0.02 \\
\hline
$9 \times 10^{-5}$ m/s & $K_p$ = 0.12, $K_i$ = 0.03, $K_d$ = 0.02 \\
\hline
$10 \times 10^{-5}$ m/s & $K_p$ = 0.15, $K_i$ = 0.02, $K_d$ = 0.01 \\
\hline
\end{tabular}
\end{table}

Looking at Figure \ref{fig:6}(c), it is clearly evident that our ART-Rx again significantly outperformed the optimal method in all scenarios. For the optimal method, there is generally an upward trend as the average flow velocity increases, with BEP values ranging between 0.28 and 0.61. Our ART-Rx remained consistently below BEP values of 0.04, with an upward trend observed starting at $8 \times 10^{-5}$ m/s. Although this may indicate that our system is starting to struggle with accurately decoding symbols as the average flow velocity increases, the low BEP results still suggest that our system is performing satisfactorily. The overall average BEP was 0.458 for the optimal method and 0.019 for the PID method.

\subsection{BEP vs. Varying Transmitter-Receiver Distance}
Next, BEP performance for various transmitter-to-receiver distances ($x_r$) was evaluated to determine the decoding accuracy of the optimal method compared to our ART-Rx. As seen in Figure \ref{fig:2}(b), the distance between Tx and Rx affects the degree to which IM disperses (denoted $w_p$) as it diffuses through the microfluidic channel \cite{Kuscu2016ModelingReceiver}. As such, various transmitter-receiver distance values between $1 \times 10^{-3}$ m and $10 \times 10^{-3}$ m were tested to evaluate the performance of ART-Rx in these scenarios. Gain scheduling was also implemented here for the nonlinear process. Obtained through Ziegler-Nichols and empirical means, Table \ref{tab:III} contains the $K_p$, $K_i$, and $K_d$ gains that were used for this set of simulations. Once again, fixed CSK concentration levels of $N_1$ = 1000 and $N_2$ = 600 were used together with numI = 700.

\begin{table}[h]
\centering
\caption{PID Gain Parameters for Transmitter-Receiver Distance Simulations}
\label{tab:III}
\begin{tabular}{|l|l|}
\hline
\textbf{Transmitter-Receiver Distance} & \textbf{PID Parameters} \\
\hline
$1 \times 10^{-3}$ m & $K_p$ = 0.18, $K_i$ = 0.02, $K_d$ = 0.005 \\
\hline
$2 \times 10^{-3}$ m & $K_p$ = 0.13, $K_i$ = 0.025, $K_d$ = 0.01 \\
\hline
$3 \times 10^{-3}$ m & $K_p$ = 0.13, $K_i$ = 0.04, $K_d$ = 0.01 \\
\hline
$4 \times 10^{-3}$ m & $K_p$ = 0.13, $K_i$ = 0.03, $K_d$ = 0.015 \\
\hline
$5 \times 10^{-3}$ m & $K_p$ = 0.12, $K_i$ = 0.045, $K_d$ = 0.005 \\
\hline
$6 \times 10^{-3}$ m & $K_p$ = 0.13, $K_i$ = 0.026, $K_d$ = 0.01 \\
\hline
$7 \times 10^{-3}$ m & $K_p$ = 0.132, $K_i$ = 0.04, $K_d$ = 0.012 \\
\hline
$8 \times 10^{-3}$ m & $K_p$ = 0.13, $K_i$ = 0.02, $K_d$ = 0.01 \\
\hline
$9 \times 10^{-3}$ m & $K_p$ = 0.16, $K_i$ = 0.02, $K_d$ = 0.008 \\
\hline
$10 \times 10^{-3}$ m & $K_p$ = 0.122, $K_i$ = 0.012, $K_d$ = 0.006 \\
\hline
\end{tabular}
\end{table}

An upward trend in BEP of both the optimal method and ART-Rx is observed in Figure \ref{fig:6}(d) as the receiver distance $x_r$ increases. This trend aligns with theoretical expectations in MC systems \cite{Kuscu2016ModelingReceiver}. As $x_r$ increases, the IM become more dispersed due to diffusion over the larger propagation distance. This dispersion results in a sparser distribution of IM and a decrease in the amplitude of the received signal $w_p$. Additionally, the longer propagation distance introduces greater delays before the IM reaches Rx. The combination of a weaker signal amplitude and increased propagation delay makes it more difficult for Rx to accurately decode transmitted information, leading to a higher BEP.

\begin{figure*}[t]
    \centering
    \subfloat[]{
    \includegraphics[width=0.32\textwidth]{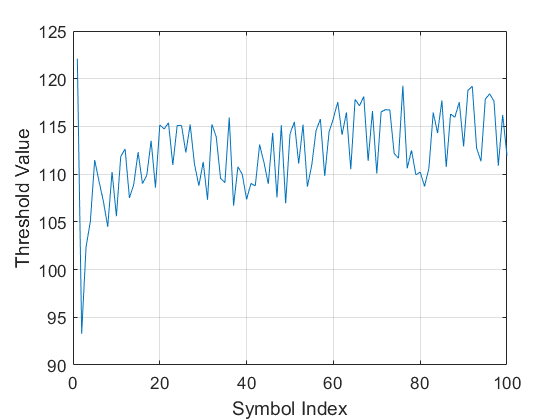}
}
\subfloat[]{
    \includegraphics[width=0.32\textwidth]{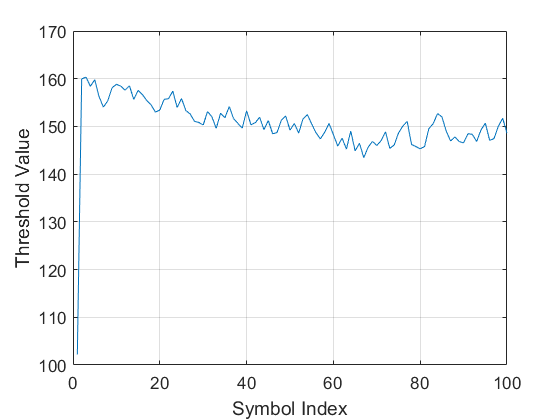}
}
\subfloat[]{
    \includegraphics[width=0.32\textwidth]{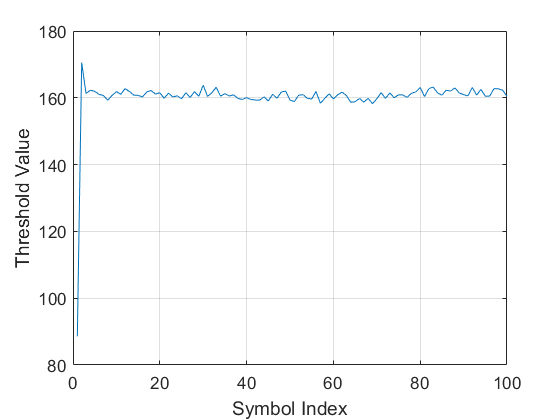}
}
    \caption{Detection threshold adjustment graphs of the PID controller when (a) $x_r$ = $1 \times 10^{-3}$, (b) $x_r$ = $5 \times 10^{-3}$, (c) $x_r$ = $10 \times 10^{-3}$. For Figures (a)-(c), the concentration levels of CSK modulation were established at $N_1$ = 1000 for bit '1' and $N_2$ = 600 for bit '0' with numI = 700.}
    \label{fig:8}
\end{figure*}

Moreover, Figure \ref{fig:8} illustrates ART-Rx's detection threshold adjustment patterns for short, medium, and long distances (\( x_r \)). In particular, fluctuations in the threshold value are more pronounced at shorter distances compared to longer ones. At \( x_r = 1 \times 10^{-3} \) meters, the threshold fluctuates significantly, ranging from approximately 105 to 120. This pronounced fluctuation is due to the higher concentration and rapid arrival of IM at shorter distances, which leads to greater variability in the received signal. At \( x_r = 5 \times 10^{-3} \) meters, the fluctuations are less pronounced, ranging between 145 and 155, reflecting increased dispersion and a smoother signal profile due to diffusion. At \( x_r = 10 \times 10^{-3} \) meters, the threshold values stabilize around 160, indicating minimal fluctuations. This stabilization occurs because the IMs experience significant dispersion and attenuation over longer distances, resulting in a weaker and more consistent received signal. Consequently, the ART-Rx adjusts its detection threshold less frequently, as the received signal varies less over time.

These data indicate that at shorter distances, the stronger and rapidly varying signal resulting from the shorter diffusion distances of the IM causes the PID controller to make more frequent and significant adjustments, leading to larger fluctuations in the detection threshold values. In contrast, as the distance between Tx and Rx increases, the signal strength attenuates, and the noise becomes more dominant. Consequently, the signal environment becomes relatively stable, and the PID controller's adjustments become less frequent and more conservative. This stabilization at longer distances suggests that the system effectively compensates for the increase in noise and the weaker signal by maintaining a consistent threshold that is less prone to fluctuation.

\subsection{BEP vs. Varying Intrinsic Diffusion Coefficient}
In this section, BEP performance was evaluated for various intrinsic diffusion coefficients ($D_0$) for the optimal method compared to ART-Rx. Various $D_0$ values between $1 \times 10^{-11}$ $m^2/s$ and $10 \times 10^{-11}$ $m^2/s$ were simulated to test the response and performance of ART-Rx against the varying environmental conditions of the channel. Obtained through Ziegler-Nichols and empirical means, Table \ref{tab:IV} contains the PID parameters $K_p$, $K_i$, and $K_d$ gains that were used for gain scheduling in this set of simulations to better conform to nonlinearities. Fixed CSK concentration levels of $N_1$ = 1000 and $N_2$ = 600 were used together with numI = 700.

\begin{table}[h]
\centering
\caption{PID Gain Parameters for Intrinsic Diffusion Coefficient Simulations}
\label{tab:IV}
\begin{tabular}{|l|l|}
\hline
\textbf{Intrinsic Diffusion Coefficient} & \textbf{PID Parameters} \\
\hline
$1 \times 10^{-11}$ $m^2/s$ & $K_p$ = 0.18, $K_i$ = 0.02, $K_d$ = 0.006 \\
\hline
$2 \times 10^{-11}$ $m^2/s$ & $K_p$ = 0.12, $K_i$ = 0.01, $K_d$ = 0.005 \\
\hline
$3 \times 10^{-11}$ $m^2/s$ & $K_p$ = 0.12, $K_i$ = 0.01, $K_d$ = 0.005 \\
\hline
$4 \times 10^{-11}$ $m^2/s$ & $K_p$ = 0.12, $K_i$ = 0.01, $K_d$ = 0.005 \\
\hline
$5 \times 10^{-11}$ $m^2/s$ & $K_p$ = 0.12, $K_i$ = 0.01, $K_d$ = 0.005 \\
\hline
$6 \times 10^{-11}$ $m^2/s$ & $K_p$ = 0.14, $K_i$ = 0.012, $K_d$ = 0.005 \\
\hline
$7 \times 10^{-11}$ $m^2/s$ & $K_p$ = 0.12, $K_i$ = 0.01, $K_d$ = 0.006 \\
\hline
$8 \times 10^{-11}$ $m^2/s$ & $K_p$ = 0.14, $K_i$ = 0.011, $K_d$ = 0.005 \\
\hline
$9 \times 10^{-11}$ $m^2/s$ & $K_p$ = 0.12, $K_i$ = 0.01, $K_d$ = 0.005 \\
\hline
$10 \times 10^{-11}$ $m^2/s$ & $K_p$ = 0.13, $K_i$ = 0.012, $K_d$ = 0.005 \\
\hline
\end{tabular}
\end{table}

Starting with the optimal method in Figure \ref{fig:6}(e), it can be observed that the BEP starts high at 0.59 for $D_0$ = $1 \times 10^{-11}$ $m^2/s$, before dramatically decreasing to 0.27 at $D_0$ = $2 \times 10^{-11}$ $m^2/s$ and 0 at $D_0$ = $3 \times 10^{-11}$ $m^2/s$, indicating a perfectly decoded transmission and outperforming our ART-Rx which had a relatively small BEP of 0.05. However, the optimal method's BEP immediately climbs rapidly at a similar rate. It increases to 0.23 at $D_0$ = $4 \times 10^{-11}$ $m^2/s$ and then 0.39 at $D_0$ = $5 \times 10^{-11}$ $m^2/s$. It then stabilizes around 0.44 throughout the rest of the simulations. The overall average BEP was 0.384 for the optimal method and 0.029 for ART-Rx.

Although the optimal method outperformed our ART-Rx at $D_0$ = $3 \times 10^{-11}$ $m^2/s$, ART-Rx still maintained a stable and low BEP throughout the simulations with BEP values no higher than 0.08, suggesting that it is rapidly adapting well to channel variations. The channel conditions may have been more ideal at $D_0$ = $3 \times 10^{-11}$ $m^2/s$ and as a result of this, the optimal method performed flawlessly. However, with the capability of ART-Rx that was consistently observed in other simulations, it is believed that performance can be matched with further tuning of the PID controller's gain scheduling.

\subsection{BEP vs. Varying Binding Rate}

Lastly, we evaluated the BEP performance of ART-Rx against the optimal method for different binding rates ($K_{b_m}$). $K_{b_m}$ values between $1 \times 10^{-17}$ $m^3/s$ and $10 \times 10^{-17}$ $m^3/s$ were simulated. This parameter was chosen to be tested with the PID controller as it directly acts on the binding properties of the receptors, contrary to the other parameters which act on the channel's environment. For a comprehensive study, it was necessary to evaluate whether the PID controller was also capable of performing optimally under such conditions. Table \ref{tab:V} contains the gains $K_p$, $K_i$, and $K_d$ that were used in this set of simulations to allow gain scheduling. The values were obtained using Ziegler-Nichols and empirical approaches. Fixed CSK concentration levels of $N_1$ = 1000 and $N_2$ = 600 were used together with numI = 700.

\begin{table}[h]
\centering
\caption{PID Gain Parameters for Binding Rates Simulations}
\label{tab:V}
\begin{tabular}{|l|l|}
\hline
\textbf{Binding Rates} & \textbf{PID Parameters} \\
\hline
$1 \times 10^{-17}$ $m^3/s$ & $K_p$ = 0.16, $K_i$ = 0.01, $K_d$ = 0.006 \\
\hline
$2 \times 10^{-17}$ $m^3/s$ & $K_p$ = 0.18, $K_i$ = 0.02, $K_d$ = 0.006 \\
\hline
$3 \times 10^{-17}$ $m^3/s$ & $K_p$ = 0.15, $K_i$ = 0.012, $K_d$ = 0.006 \\
\hline
$4 \times 10^{-17}$ $m^3/s$ & $K_p$ = 0.16, $K_i$ = 0.011, $K_d$ = 0.005 \\
\hline
$5 \times 10^{-17}$ $m^3/s$ & $K_p$ = 0.12, $K_i$ = 0.01, $K_d$ = 0.005 \\
\hline
$6 \times 10^{-17}$ $m^3/s$ & $K_p$ = 0.14, $K_i$ = 0.0115, $K_d$ = 0.005 \\
\hline
$7 \times 10^{-17}$ $m^3/s$ & $K_p$ = 0.22, $K_i$ = 0.014, $K_d$ = 0.006 \\
\hline
$8 \times 10^{-17}$ $m^3/s$ & $K_p$ = 0.146, $K_i$ = 0.011, $K_d$ = 0.007 \\
\hline
$9 \times 10^{-17}$ $m^3/s$ & $K_p$ = 0.16, $K_i$ = 0.012, $K_d$ = 0.006 \\
\hline
$10 \times 10^{-17}$ $m^3/s$ & $K_p$ = 0.163, $K_i$ = 0.011, $K_d$ = 0.007 \\
\hline
\end{tabular}
\end{table}

Observing the results in Figure \ref{fig:6}(f), the wide performance gap is again immediately noticeable between the optimal method and ART-Rx. The optimal method had high BEP values between 0.33 and 0.6, with a generally upwards trend as $K_{b_m}$ increases. ART-Rx had consistently low and stable BEP values in all simulated $K_{b_m}$ values. All BEP results were below 0.06 with a perfectly decoded transmission observed at $K_{b_m}$ = $4 \times 10^{-17}$ $m^3/s$. The overall average BEP was 0.513 for the optimal method and 0.024 for ART-Rx. This again suggests that ART-Rx performed optimally throughout the course of this set of simulations with varying $K_{b_m}$.

\subsection{Performance Evaluation of ART-Rx}

Table \ref{tab:VI} summarizes the BEP results obtained from the simulations performed.

\begin{table}[h]
\centering
\caption{Summary of BEP Results}
\label{tab:VI}
\begin{tabular}{|l|l|l|}
\hline
\textbf{Simulation} & \textbf{Optimal Method BEP} & \textbf{ART-Rx BEP} \\
\hline
numI & 0.33 & 0.018 \\
\hline
$N_2$ & 0.259 & 0.067 \\
\hline
u & 0.458 & 0.019 \\
\hline
$x_r$ & 0.485 & 0.074 \\
\hline
$D_0$ & 0.384 & 0.029 \\
\hline
$K_{b_m}$ & 0.513 & 0.024 \\
\hline
\end{tabular}
\end{table}

As evident in Table \ref{tab:VI}, the ART-Rx system consistently outperformed the statistically based optimal detection threshold method in all simulations. ART-Rx achieved significantly lower BEP values, demonstrating its robustness and effectiveness in various channel and noise conditions.

\paragraph{Comparative Analysis with ML Methods}

Compared to existing adaptive thresholding techniques based on ML techniques ~\cite{Debus2023ReinforcementMobility, Lee2017MachineCommunications, Qian2018ReceiverNetworks, Qian2019MolecularOptimization, Jing2024PerformanceCommunication}, ART-Rx offers several advantages:

\begin{itemize}
    \item \textbf{Computational Efficiency:} The implementation of a PID controller is less computationally intensive than obtaining data, training, and deploying ML models, which makes it suitable for resource-constrained nanoscale devices.
    \item \textbf{Simplicity:} ART-Rx does not require large data sets for training or complex algorithms, simplifying implementation and reducing the risk of overfitting. Having a simple system also introduces redundancy for any potential failures by making diagnosis processes easier and provides opportunities for modular expansions to further improve the system.
    \item \textbf{Real-Time Adaptation:} ART-Rx can adjust the biorecognition unit's detection threshold in real-time without the need for retraining, which is beneficial in dynamically changing environments.
\end{itemize}

However, considering how PID controllers were originally designed for linear use cases, ML-based methods may offer better performance in highly nonlinear or unpredictable environments if sufficient training data are available. Incorporating elements from these methods and introducing hybrid solutions could potentially further enhance ART-Rx

\paragraph{Impact of PID Parameter Tuning}

The superior performance of the ART-Rx system is attributed to its ability to dynamically and rapidly adjust the detection threshold in real-time. The effectiveness of the PID controller is highly sensitive to the tuning of the gains $K_p$, $K_i$, and $K_d$. Proper tuning of these parameters is critical to achieve optimal performance under varying microfluidic channel conditions.

We initially applied the Ziegler-Nichols method \cite{Ziegler1993OptimumControllers} to set the starting values for these gains, followed by iterative simulations to fine-tune the parameters before running various simulation sets to evaluate the performance of ART-Rx. This approach balanced system responsiveness and stability to achieve optimal performance. Each of the gain values used in the simulations were obtained using this approach.

\paragraph{Sensitivity Analysis of PID Gains}

Each PID gain parameter plays a distinct role in the controller's performance \cite{Wang2020BasicsControl, KiamHeongAng2005PIDTechnology}:

\textbf{Proportional Gain ($K_p$):} Influences the system's speed in adjusting the detection threshold. Higher $K_p$ can lead to faster responses, but can cause overshooting and instability.

\textbf{Integral Gain ($K_i$):} Eliminates steady-state errors by accounting for cumulative error. Excessive $K_i$ can lead to slow responses and integral wind-up.

\textbf{Derivative Gain ($K_d$):} Provides predictive control by counteracting the rate of change of the error signal. Essential for mitigating sudden changes in noise levels.

Fine-tuning these parameters allowed ART-Rx to consistently outperform the optimal method with significantly smaller BEP and BER metrics. Gain scheduling was also used, with different sets of PID gains applied for different channel conditions, which helped the controller to better adapt to the channel's nonlinear conditions and improve system performance. Advanced PID tuning techniques, such as auto-tuning, could further enhance performance under unknown and dynamic channel conditions \cite{Sukede2015AutoController, Xin2020ResearchAuto-tuning}.

\subsection{Practical Implementation Challenges}

Although ART-Rx shows promise in simulations, several practical challenges must be addressed for real-world implementation.

\subsubsection{Hardware Limitations}

Implementing ART-Rx on a nanoscale presents hardware and fabrication challenges. The integration of all components into an SoC requires careful consideration of power consumption, processing capabilities, and physical space constraints. Developing energy-efficient hardware architectures and leveraging emerging nanotechnology fabrication techniques will be crucial to the realization of ART-Rx on the nanoscale.

\subsubsection{Computational Resources}

Although the PID controller is computationally efficient compared to ML methods, real-time processing in a nanoscale device may still be demanding. Optimizing the algorithm for low-power operations and exploring hardware acceleration, such as using specialized processing units, could alleviate this issue.

\subsubsection{Network Integration}

Integrating the ART-Rx system with existing MC networks and devices requires compatibility with communication protocols and standards. Determining a universal standard communication protocol for seamless interoperability and security is of utmost importance prior to system deployment.

\subsection{Limitations and Future Work}

\subsubsection{Limitations of PID Controllers in Nonlinear Systems}

PID controllers are linear control mechanisms, and MC channels are highly nonlinear and varying in time. The PID controller may not adequately compensate for complex nonlinear behaviors in real-world scenarios, leading to sub-optimal performance in practical applications.

Future work should explore advanced control strategies better suited for nonlinear systems, such as but not limited to:

\begin{itemize}
    \item \textbf{Adaptive Control:} Adjusts controller parameters in real-time based on system behavior.
    \item \textbf{Nonlinear Control Methods:} Utilizes control laws designed for nonlinear dynamics \cite{YaolongTanAdvancedSystems, Pettit1997SimpleSystems}.
    \item \textbf{Model Predictive Control (MPC):} Uses models to predict future system behavior and optimize control actions \cite{Bemporad2006ModelTools}.
\end{itemize}

\subsubsection{Simplification of the Noise Model}

Our simulations used a simplified noise model that focuses on the use of interferer molecules as noise in the Smoldyn simulator \cite{Andrews2012SpatialSimulator}, which may not fully capture the complexities of microfluidic channels in the real world. Additional noise sources, such as \cite{Kuscu2019TransmitterTechniques, Shih2012ChannelCommunications}:

\begin{itemize}
    \item Variable flow-induced turbulence
    \item Multi-path diffusion effects
    \item Molecular degradation
    \item Environmental fluctuations (e.g., temperature, pH levels)
    \item Non-Gaussian noise distributions
    \item Molecular interactions (e.g., binding competition)
\end{itemize}

should be considered in future models. Incorporating these factors will provide a more accurate assessment of the performance of the ART-Rx system in the real world and will further enhance its robustness.

\subsubsection{Future Research Directions}

To address the limitations and challenges, future research should focus on the following:

\begin{itemize}
    \item \textbf{Developing Advanced Control Strategies:} Implement controllers suitable for nonlinear and time-varying systems.
    \item \textbf{Enhancing Noise Models:} Incorporate complex noise sources and environmental factors into simulations.
    \item \textbf{Experimental Validation:} Conduct laboratory experiments on testbeds to validate simulation results under realistic conditions.
    \item \textbf{Optimizing Hardware Implementation:} Design energy-efficient hardware architectures for the PID controller and the ART-Rx system as a SoC.
    \item \textbf{Exploring Hybrid Approaches:} Combine the simplicity of PID control with ML techniques to improve adaptability.
\end{itemize}

\subsection{Potential Applications and Implications}

The ability of the ART-Rx system and PID controllers to maintain low BER and BEP under varying channel conditions positions control theory as a promising solution to improve the reliability of data transmission in practical and nanoscale MC systems. Its simplicity and computational efficiency make it suitable for integration into nanoscale devices with limited processing capabilities.

Potential applications of the ART-Rx system include:

\begin{itemize}
    \item \textbf{Healthcare Monitoring:} Robust and reliable communication in implantable medical devices and biosensors for continuous health monitoring \cite{Kuscu2016OnBiosensors, Akyildiz2008Nanonetworks:Paradigm, Veletic2020ModelingDelivery}.
    \item \textbf{Brain-Machine Interfaces (BMI):} High-fidelity signal transmission in neural interfaces for advanced prosthetics and various other cases of medical and non-medical use \cite{Nakano2014MolecularIssues, Nakano2012MolecularChallenges}.
    \item \textbf{Internet of Bio-Nano Things (IoBNT):} Robust communication between nanoscale device networks in complex biological environments \cite{MuratKuscu2021InternetChallenges, Akyildiz2015TheThings}.
    \item \textbf{Environmental Sensing:} Deployment in environmental monitoring systems to detect pollutants or biochemical agents with high sensitivity \cite{Nakano2019ApplicationsSystems, Ahmed2022MolecularSciences}.
    \item \textbf{Organoid Intelligence:} Enabling advanced communication interfaces with brain organoids to support the development of organoid intelligence and integration with external devices \cite{Smirnova2023OrganoidIntelligence-in-a-dish}.
\end{itemize}

By enhancing signal detection in MC systems, ART-Rx has the potential to contribute significantly to the advancement of nanotechnology and its applications in various fields.

\section{Conclusion}

In this paper, we proposed ART-Rx, a novel adaptive threshold receiver based on PID control, to improve signal detection in diffusion-based microfluidic MC systems. ART-Rx adjusts the detection threshold in discrete real-time, adapting to dynamic channel and noise conditions while maintaining robustness and computational efficiency. We also introduced a SoC design that integrates ART-Rx.

Simulations demonstrated that ART-Rx significantly outperforms the statistically-based optimal detection threshold method under challenging scenarios, including high interference levels, closely spaced CSK levels, and extreme channel conditions. ART-Rx maintains consistently low BER and BEP, highlighting its potential to improve the reliability of MC systems in dynamic environments.

Despite promising results, limitations include the applicability of PID controllers in nonlinear, time-varying systems and the simplification of the noise model. Future research should explore advanced control strategies suited for nonlinear dynamics, incorporate more sophisticated noise models, optimize PID parameter tuning, investigate hybrid approaches with ML, and address practical implementation challenges like SoC hardware limitations and system integration in the context of MC.

In summary, ART-Rx represents a significant advancement in MC Rx design, offering a simple, robust, and effective approach to improving signal detection in nanoscale communication systems. By addressing these limitations through future research, ART-Rx and control theory can substantially contribute to the development of reliable MC systems with potential applications in healthcare monitoring, organoid intelligence, BMI, environmental sensing, and the realization of IoBNT and IoE.

\bibliographystyle{ieeetr}
\bibliography{references}

\end{document}